\documentclass[singlecolumn,epjc3]{svjour3}

\smartqed  

\RequirePackage{graphicx}

\usepackage{amsmath,amssymb}
\usepackage{lineno,hyperref}

\journalname{Eur. Phys. J. C}

\begin{document}

\title{Solar system tests in constraining parameters of dyon black holes}

\author{Farook Rahaman\thanksref{e1,addr1}
\and Sabiruddin Molla\thanksref{e2,addr2} \and Amna Ali\thanksref{e3,addr3} \and Saibal Ray\thanksref{e4,addr4}. }

\thankstext{e1}{e-mail: rahaman@associates.iucaa.in}
\thankstext{e2}{e-mail: sabiruddinmolla111@gmail.com}
\thankstext{e3}{e-mail: amnaalig@gmail.com}
\thankstext{e4}{e-mail: saibal@associates.iucaa.in}

\institute{Department of Mathematics, Jadavpur University, Kolkata
700032, West Bengal, India\label{addr1}
\and
Department of Mathematics, Jadavpur University, Kolkata
700032, West Bengal, India\label{addr2}
\and
Department of Mathematics, Jadavpur University, Kolkata
700032, West Bengal, India\label{addr3}
\and
Department of Physics, Government College of Engineering and Ceramic Technology, Kolkata
700010, West Bengal, India \& Department of Natural Sciences, Maulana Abul Kalam Azad University of Technology, 
Haringhata 741249, West Bengal, India\label{addr4} }

\date{Received: date / Accepted: date}

\maketitle

\begin{abstract}
In the present paper we examine the possibility of  constraining dyon black holes based on the available observational data at the scale of the Solar system. For this we consider the classical tests of general relativity, viz., the perihelion precession of  the  planet  Mercury and the deflection of light by the Sun. In connection to mathematical analysis we are considering static and spherically symmetric dyon black hole which carries both the electric and magnetic charge simultaneously, which are encoded it by the parameters $\lambda_0$ and $\beta_0$. We
constrain these two parameters using the Solar system tests and obtain the permissible range from theoretical analysis based on our model and later on compare them with the available observational data.
\end{abstract}


\section{Introduction}
Black holes are believed to be the most promising candidates to address the quantum effects of gravity, as they
represent in many ways the possible testing ground for string theory~\cite{Iorio2015,Debono2016,Vishwakarma2016}. 
The reason behind this is that, in order
to probe gravity beyond general relativity (where classical physics breaks down), one needs a
singularity, which can be found in the center of a black hole.  Although mathematical formulations have
been done for several formal aspects of string theory (quantum gravity), the subject still remains enigmatic
as well as remote from other areas of physics as far as measurements or observations are concerned. 
Sub-field of string theory or string phenomenology otherwise known is well
developed that pertains to resolve observational issues of string theory. In the present day scenarios string
phenomenology has gained potential to unravel particle physics related as well as cosmological observations in a more deeper sense.

Earlier black hole  were considered merely a mathematical
entity rather than real existing astrophysical objects of the universe. Recently, especially after the successful
detection of gravitational waves from black holes collisions~\cite{Abbott2016,Cervantes-Cota2016} and the possibility of looking at an
astrophysical black holes~\cite{Ricarte2015}, black holes have become a hot topic in the current physics research~\cite{Cembranos2015,Olmo2015,Batic2016,Oikonomou2016,Silva2016,Barcelo2016,Rodriguez2017,Chakraborty2017,Nova2017,Kawai2017,Malafarina2017,Erbin2017,Czinner2017,Prudencio2017,Gray2018,Iqbal2018,Kruglov2018,Tretyakova2018}.
Black holes arise from low energy effective string theories as a result of various compactification
schemes. String compactification to four dimensions involves a manifold and generates multiple scalar fields.
For example, many black hole solutions have been obtained in the
low-energy string theories in which the Kalb-Ramond field, dilaton field and
gauge field are incorporated~\cite{Bowick1988,Gibbons1988,Campbell1990,Campbell1991,Hsu1991,Garfinkle1991,Lee1991,Horne1992}.
 The massless scalar fields coupled with the vector fields, which also naturally occur
 in the bosonic sector of such four-dimensional effective theories, in addition to gravity and scalar fields.

The Einstein-Maxwell-axion-dilaton system includes all the main features of low-energy string theory, where
the two scalar fields couple to the abelion vector fields. The dilaton couples to the square of the
Maxwellian field strength $F^2$, whereas the axion couples to the topological term $F*F$ 
and is therefore a pseudoscalar. The presence of these scalar fields
affect the basic properties of the black hole. The set of static spherically symmetric solutions of the
Einstein-Maxwell-dilaton system, falls in the class of dilaton black holes.
They possess either purely a magnetic or electric Maxwell field and have a vanishing axion~\cite{Gibbons1988,Garfinkle1991}. 
However, when gravity is coupled to dilaton and gauge field there appears a new axion field and the solutions become dyonic,
where the black holes carry electric and magnetic charge simultaneously~\cite{Lee1991,Shapere1991}.

Now, to be a viable black hole solution, the proposed model need to pass the observational tests of astrophysics. 
One of the possibilities of such observation-based testing black hole model at an astrophysical scale
are the solar system tests. The solar system tests are the classical tests of general relativity, which
have been analyzed for various gravitational theories with large non-compactified higher-dimensions~\cite{Liu2000}, 
by using an analogue of the four-dimensional Schwarzschild metric and have imposed strong constraints on it.
This paper however concerns with the solar system tests, namely the perihelion precession and bending of light for a 
static spherically symmetric dyonic black hole. 

Therefore, the paper basically contains a technical result in gravity which treats dyonic black holes. We refer to some references regarding dyonic black holes in string theory and connect the idea with the electromagnetism expecting the possible existence of astrophysical charged black holes. It is to note that electromagnetic black holes, especially the Reissner-Nordstr{\"o}m black holes, have been studied widely in literature~\cite{Iorio2012}. However, our paper emphasizes that if such an object do exists, then what range of its parameters are viable to satisfy the observations.

\section{Background formalism for dyonic black holes}
In the early nineties, Cheng et al.~\cite{Cheng1994} obtained an exact
solution of the low-energy string theory representing static
spherically symmetric dyonic black hole. They have chosen the 
four-dimensional effective string action in which gravity is coupled to
dilaton and electromagnetic field given by 
\begin{equation} 
I =  \int d^4x\sqrt{-g}[-R + 2(\nabla{ \phi_0})^2  + e^{-2{\phi_0}} F^2]. \label{action}
\end{equation}

We consider here a static and spherically symmetric configuration given by
\begin{equation}
ds^2=-A(r) c^2 dt^2 +B(r) dr^2+C(r)(d\theta^2+\sin^2\theta d\phi^2),
\end{equation}
where
\begin{equation}
A = \frac{1}{c^{2}}\left(1- \frac{2M}{r^2} \sqrt{r^2 + {\lambda}^2} +
\frac{\beta}{r^2}\right),
\end{equation}

\begin{equation}
B = \frac{r^{2}}{r^{2}+\lambda^{2}}\left(\frac{1}{1- \frac{2M}{r^2} \sqrt{r^2 + {\lambda}^2} +
\frac{\beta}{r^2}}\right),
\end{equation}

\begin{equation}
C=r^{2},
\end{equation}

with $\lambda = ({Q_e}^2 e^{2{\phi}_0} - {Q_m}^2 e^{2{\phi}_0})/2M$ and $\beta = ({Q_e}^2 e^{2{\phi}_0} + {Q_m}^2 e^{2{\phi}_0}).$

The electrically (or magnetically) charged dilaton black holes are the special case of the dyonic black holes. 
However, it is to be noted that unlike the electric or magnetic charged dilaton holes, the Hawking temperature of the dyonic
black hole depends on both the electric ($Q_e$) as well as magnetic ($Q_m$) charges and vanishes as they tend to extremal values. 
Except $Q_e$ and $Q_m$ the dyonic black hole solution is characterized by the other two parameters, viz., mass $(M)$ and asymptotic value of the scalar dilaton $({\phi}_0)$.

In the following text we perform the solar system tests on these solutions which are the classical tests of general relativity.

\section{Solar system tests for dyonic black holes} 
The perihelion precession of the Mercury and deflection of light by the Sun are the 
fundamental tests at the level of the Solar system. These classical tests of general relativity
have been used successfully to test the Schwarzschild solutions and some of its generalization. 
In order to perform the above mentioned tests, we consider 
the geometry outside of the Sun which is a compact stellar type object comprising of 
a specific static and spherically symmetric vacuum solutions in the context of dyonic black hole.

\subsection{Perihelion precession} The eight major planets of the Solar system are 
rotating around the Sun in elliptical orbits which are approximately 
coplanar with each other. The point on the orbit of a planet at which it is
closest to the Sun is known as {\it perihelion}. The perihelion of a
particular planet remains fixed in space when the relatively weak
interplanetary gravitational interaction is neglected. However, when this
interaction is considered the perihelion slowly precesses
around the Sun. This perihelion precession of the
planets has been treated as one of the most important tests to check
the correctness of general relativity.

\begin{table}
\centering 
\caption{Perihilion precession for different parametric values} \label{table1}
\begin{tabular}{@{}lrrrrrrrrr@{}}
\hline 
$\lambda_0$  &~~$\beta_0$  &~~$u_0/b^2$  &~~$\Delta \phi $ \\
                   &                   &                     & (arcsec/cen)\\             
\hline 
0.009 & 0.1 & 1.007 & 43.173\\
\hline 
0.01 & 0.1 & 1.001 & 43.657\\
\hline 
0.1   &-0.1 & 1.005 & 42.38\\
\hline 
0.1 & 0.2 & 0.998 & 43.672\\
\hline 
1 & 2 & 1.001 & 71.60\\
\hline 
1 & 10 & 0.99 & 128.887\\
\hline 
\end{tabular}
\end{table}

Now, we begin with the Lagrangian which can be written as
\begin{equation}
 \mathcal{L}=-A c^2 {\dot t}^2 + B{\dot r}^2 + C{\dot{\theta}}^2 + C \sin^2{\theta}{\dot{\phi}}^2,
\end{equation}
where dot over any parameter implies differentiation with respect to the affine parameter $ {\tau}$.

It is known that the gravitational field is isotropic and hence there is conservation of angular momentum. So geodesics of the
particles (either massive planets or massless photons) are planar. Without loss of generality, we can choose our coordinates 
in such a way that this plane is the equatorial plane given by keeping fixed ${\theta}= {\pi}/2$. 

Therefore, the Lagrangian takes the form
\begin{equation}
\mathcal{L}=-A c^2 {\dot t}^2 + B{\dot r}^2 + C {\dot{\phi}}^2, \label{lagrangian}
\end{equation}
with light-like particle photon, $\mathcal{L} = 0 $ and for any time-like particle, $\mathcal{L}=1$.

Considering the generalized coordinates $q_i$  and generalized
velocities ${\dot q}_i$, the Euler-Lagrange equations become
\begin {equation}
\frac{d}{ds}\left(\frac{\partial \mathcal{L}}{\partial{\dot q}_i}\right)- \frac{\partial \mathcal{L}}{\partial {q}_i}=0,
\label{EL}
\end {equation}
given that $ Ac^2 {\dot t}= E $ and $ C \dot{\phi}= L$,  where $E$ and 
$L$ are the energy and momentum of the particle respectively, such that 
$ \dot t =E/c^2 A $ and $ \dot{\phi}= L/C$.

Using  the above notations, from Eqs. (\ref{lagrangian}) and (\ref{EL}) we get~\cite{Bowick1988}
\begin{equation}
\left(\frac{dr}{d{\phi}}\right)^2 + \frac{C}{B} = \frac{\mathcal{L}C^2}{B L^2} + \frac{E^2 C^2}{ABL^2 c^2}.
\end{equation}

Again by substituting $r = 1/U$ in the above equation, one can write
\begin {equation}
\left(\frac{dU}{d \phi}\right)^2 + \frac{CU^4}{B} = \frac{\mathcal{L} C^2 U^4}{BL^2} + \frac{E^2 C^2 U^4}{ABL^2 c^2}.
\end{equation}

Considering the following transformations $1/B = 1- f(U) $ and  $C= [1/U^2] + [g(U)/U^4]$ the above equation can be written as
\begin {eqnarray}
\left(\frac{dU}{d\phi}\right)^2 + U^2 = U^2 f(U) + f(U)g(U) - g(U) \nonumber\\
+  \frac{\mathcal{L} C^2 U^4}{BL^2}+\frac{E^2 C^2 U^4}{ABL^2 c^2}\equiv G(U). \label{G}
\end{eqnarray}

Differentiating with respect to $U$, one gets
\begin {equation}
\frac{d^2 U}{d{\phi}^2} + U = F(U),
\end{equation}
where
\begin{equation}
F(U) = \frac{1}{2} \frac{dG(U)}{dU}. \label{F}
\end{equation}

Since the planetary orbits are nearly circular, so the circular orbit  $U= U_0$  can be obtained from the equation
\begin {equation}
F(U_0)= U_0.
\end{equation}

Now, for the deviation from the circular orbit the perihelion precession is given by
\begin{equation}
\sigma = \frac{1}{2}\left(\frac{dF}{dU}\right)_{U = U_0}.
\end{equation}

Therefore, for complete rotation, the advancement of perihelion becomes
\begin{equation}
\Delta \phi = \phi - 2\pi \thickapprox 2\pi \sigma. \label{pp}
\end{equation}

For dyonic black hole the parameters can be written as
\begin{eqnarray}
G(U)= U^2 + (1 + U^2 \lambda^2) \left(\frac{1}{L^2} - U^2\right) \nonumber \\ \left(1 -  2 M U \sqrt{1 + U^2 \lambda^2}+ \beta U^2\right) \nonumber \\ +\frac{E^2}{L^2}\left(1 + {\lambda^2}{U^2}\right),
\end{eqnarray}

\begin{eqnarray}
&\qquad\hspace{-9cm} F(U)= [U\sqrt{1+U^2 L^2}(\beta-2L^2 U^2 \beta \nonumber \\ 
&\qquad\hspace{-6cm} +(1+E^2+U^22\beta-L^2(2+3U^2\beta)))\lambda^2 \nonumber \\ + M(1+U^2\lambda^2)(-1+U^2 \nonumber \\ 
(-4\lambda^2+L^2(3+6U^2\lambda^2)))]/L^2 \sqrt{1+U^2 \lambda^2}.
\end{eqnarray}

In order to find the circular orbits, we represent the parameters as follows: $\lambda = \lambda_ 0 M$,
 $\beta = \beta_ 0 M^2 $ and $U_0 = x_0 /M,$ where $\lambda_0$, $\beta_0$
and $ x _0 $ are dimensionless parameters, respectively and find the roots of the equation
 $F (U_ 0 ) = U_0$, which can be written as
\begin{eqnarray}
&\qquad\hspace{-9cm} 3x_0^2-b^2(4x_0^2 \lambda_0^2+1)=-6x_0^4\lambda_0^2+\nonumber\\ 
\frac{1}{\sqrt{1+x_0^2\lambda^2}} [3x_0^5\beta_0\lambda_0^2 -2x_0^3\{-\beta+(-1+b^2\beta_0)\lambda_0^2\}+ \nonumber\\ x_0(1+b^2\{\beta_0+(1+E)\lambda_0^2\})],\nonumber\\ \label{root}
\end{eqnarray}
where $b^2 = M^2 /L^2 $.

To obtain the perihelion precession, one needs to know the value of the parameter $L$ in terms of
the orbit parameters. According to Harko et al.~\cite{Harko2011}
\begin{equation}
\frac{1}{L^2} = \frac{c^2}{GMa(1- e^2)},
\label{l}
\end{equation}
where $a$ is the semi-major axis and $e$  is the eccentricity of the orbit.

For the planet Mercury we have $a=57.91 \times 10^{11}$~cm,~$e =0.205615$, while
the Solar mass $M= M_\odot = 1.989 \times 10^{33}$~gm,
~$c=2.998 \times 10^{10}$~cm/s,~$G = 6.67 \times 10^{-8}$~\text{$cm^3/gm~s^2$}~\cite{Shapiro1971,Shapiro1976}.
The Mercury also completes $415.2$ revolutions in each century. By the use of these numerical values we first
obtain $b^2 = M/a (1 - e^ 2) = 2.66136 \times 10^{−8}$. Hence putting $E=1$ we perform 
a first order series expansion of the square root in Eq. (\ref{root}) and obtain the standard general relativistic
equation $3x_0^2 - x _0 + b^ 2 = 0$, with the physical solution $x_0^{GR}\approx b^ 2$. 
The value of $x_0 $ is obtained by solving the nonlinear algebraic equation Eq. (\ref{root}) numerically which depends
on the parameters  $\lambda_0$ and $\beta_0$. In general this equation has many roots for different parametric values of $\lambda_0$
and $\beta_0$, but we shall concentrate only on the positive values of $x_0$ which are close to $x_0^{GR}\approx b^2$.
Using these numerical values of $x_0$ we evaluate the perihelion precession angle from Eq. (\ref{pp}). The angle of perihelion precession
 $\Delta \phi $ for different values of $\lambda_0$ and $\beta_0$ are shown in Table 1. We note that 
for some values of parameters the precession is well within the observed value of the perihelion precession of the
planet Mercury which is $\Delta \phi_{obs}=43.11 \pm 0.21~$arcsec/cen ~\cite{Bohmer2010,Roy2015}.

\subsection{Deflection of light} 
Henry Cavendish in 1784 (in an unpublished manuscript) and J.G. Von Soldner in 1804
showed the deviation of light rays near a  massive object by using Newtonian gravity. In the early twentieth century 
Einstein calculated the deflection of light by his general relativity, and this came as twice the Newtonian value. 
Later on, Arthur Eddington and his collaborators verified this phenomenon by observing the change in 
position of stars during a total solar eclipse in May 1919. So this deflection of light is 
also one of the observational verification of general relativity.

\begin{table}
\centering 
\caption{Angle of deflection for different parametric values}
\label{table2}
\begin{tabular}{@{}lrrrrrrrrr@{}}
\hline 
$\lambda_0$  & $\beta_0$  & $ C_1$  & $\delta\phi $ \\
 \hline
0.009 & 0.1 & 0.48 & 1.78\\
\hline
0.01 & 0.1 & 0.59 & 1.753\\
\hline
0.013 & 0.2 & 1 & 1.79\\
\hline
0.011 & 0.1 & 1 & 1.70\\
\hline
0.1   &-0.1 & 60.1 & 1.752\\
\hline
0.1 & 0.2 & 60.1 & 1.75\\
\hline
1 & 2 & 5938 & 1.75\\
\hline
\end{tabular}
\end{table}

To investigate the bending of light in the vicinity of dyon black hole we start
with Eq. (\ref{lagrangian}) by assuming $\mathcal{L} = 0 $ which now reads
\begin{eqnarray}
\left(\frac{dU}{d{\phi}}\right)^2 + U^2 = U^2 f(U) +f(U)g(U)\nonumber\\
 - g(U) +\frac{E^2 C^2 U^4}{ABL^2 c^2} \equiv P(U).
\end{eqnarray}

Differentiating with respect to $U$, one gets
\begin{equation}
\frac{d^2 U}{d{\phi}^2} + U = Q(U),
\end{equation}
where
\begin{equation}
Q(U)= \frac{1}{2}\frac{dP(U)}{dU}.
\end{equation}

Now, we solve the equation by successive approximation, starting from
the straight line (path without gravitating body) as zeroth
approximation such that $ U= \cos \phi/R_0$, where $ \phi = 0 $
is the point of nearest approach to the Sun's surface. Ideally, $
R_0 $ would be the Solar radius. Substituting this on the right hand
side of Eq. (22) for $U$, we get
\begin{equation}
\frac{d^2 U}{d{\phi}^2} + U = Q\left(\frac{\cos\phi}{R_0}\right).
\end{equation}

This gives the general solution for $U = U(\phi)$. The light ray from infinity comes at the asymptotic angle 
$\phi =-(\pi/2+\epsilon)$ and goes out to infinity at an asymptotic
angle $\phi =\pi/2+\epsilon$. The angle $\epsilon $ is obtained as a solution
of the equation $U (\pi/2+\epsilon) = 0$. The total deflection
angle of the light ray is given by $ \delta=2\epsilon $. In dyonic black hole scenario,
keeping terms up to $(1/R_0)^6 $, above equation takes the following form
\begin{eqnarray}
& & \frac{d^2 U}{d{\phi}^2} + U =\frac{21}{8}\lambda_0^{4}\left(\frac{\cos \phi}{r_0}\right)^6-3\beta_0\lambda_0^{2}\left(\frac{\cos \phi}{r_0}\right)^5\nonumber\\
& &+\frac{15}{2}\lambda_0^{2}\left(\frac{\cos \phi}{r_0}\right)^4-2\left(\beta_0+\lambda_0^{2}\right)\left(\frac{\cos \phi}{r_0}\right)^3\nonumber\\
& &+3\left(\frac{\cos \phi}{r_0}\right)^2+\frac{c^{2}}{r_0^{2}}\lambda_0^{2}\left(\frac{\cos \phi}{r_0}\right),
\end{eqnarray}
where $L= E R_0/c$, $E=1$ and representing all the variables in the dimensionless form such that $r_0=R_0/ M$, $\lambda_0 = \lambda/ M $ and $\beta_0 = \beta/ M^2$.

We solve this equation considering the first approximation and substitute $\phi = \pi/2 + \epsilon$, $ U = 0$ and 
use the relations $\cos(\pi/2 + \epsilon) = -\sin \epsilon$, $\cos (\pi + 2\epsilon) = -\cos 2\epsilon $, $\sin\epsilon \approx \epsilon$ 
and $\cos 2\epsilon\approx 1$. Eventually, for $\epsilon$ we get
\begin{eqnarray}
&\qquad\hspace{-11cm} \epsilon=[(40 R_0^3 (16 R_0 - 3 \pi \beta_0) \nonumber \\ +
 10 R_0 (4 R_0 (32 + (-3 + 2 c^2 M^2) \pi R_0) \nonumber \\+
    15 \pi \beta_0) \lambda_0^2 +
 384 \lambda_0^4)]/[(5 R_0 (64 C_1 R_0^4 - 7 \beta_0 \lambda_0^2 \nonumber \\+
   8 R_0^2 (\beta_0 + \lambda_0^2 - 2 c^2 M^2 \lambda_0^2)))].
\end{eqnarray}

Here $C_1$ is the arbitrary constant. For the Sun, by taking $R_0$ = $R_\odot  = 6.955  \times 10^{10}$~cm,
where $R_\odot$  is the radius of the Sun, one can obtain the value of $r_0$ as $r_0 = 4.71194 \times 10^5$. 

We evaluate the angle of light deflection $\delta\phi=2\epsilon$
for different values of parameter $\lambda_0$ and $\beta_0$ in Table 2. It is to be noted that for various parameters 
the theoretical values of angle of deflections are in agreement with the observational value obtained
 from long baseline radio interferometry~\cite{Robertson1991,Lebach1995},
which gives $\delta\phi_{LD} = \delta\phi_{LD}^{(GR)}(1 + \Delta_{LD})$, with $\Delta_{LD} \leq
0.0002 \pm 0.0008$, where $\delta\phi^{(GR)}_{LD} = 1.7510$~arcsec.

\section{Discussion and conclusion}
The classical tests of general relativity provide a very powerful tool for constraining the allowed
parameter space of various astrophysical solutions, such as the black hole and brane world solutions.
It provides a deeper insight into the physical nature and properties
of the corresponding astrophysical object and space-time metric. In the present work, we have analyzed
the dyonic black hole, which carries both electric and magnetic charge simultaneously. The black holes have been 
considered so far as hypothetical as well as intrigued objects which have sparked interest due to recent 
detection of the gravitational waves formed by the merger of black holes.

 In this context, the classical tests of general relativity, namely the perihelion precession and deflection
of light are considered for the specific static and spherically symmetric solutions
of dyonic black hole. This black hole has two free parameters, $\lambda_0$ and $\beta_0$,
which quantify the electric and magnetic charge in it. We have  constrained the parameters 
by the perihelion precession of the planet Mercury and found $0.01 \geq \lambda_0 \geq 0.009$ 
and $\beta_0=0.1$ which agree satisfactorily to the observational result. However, from the study of the 
deflection of light by the Sun it is noticed that the admissible range of the parameters can be extended
as $0.1 \geq \lambda_0 \geq 0.009$ and $0.02 \leq \beta_0 \leq 0.1$ (vide Table 2).

At this juncture we would like to emphasis that a crucial issue of the present
study consists of the perihelion precession. Indeed, the theoretically predicted 
values of Table 1 can be compared to the most recent experimental determinations
of the anomalous perihelion precessions for all the inner planets along with the
Saturn by the team producing the planetary ephemerides. It must be stressed here
that, actually, there are no non-zero anomalous perihelion precessions at a 
statistically significant level since all are statistically compatible with zero. Indeed, the
best values are smaller than the errors released. Thus, such values can be used to
infer upper bounds on the parameters of the perihelion precessions as shown in
Table 1 by comparing them with such recent experimentally determined values.
It has been done in a number of papers in the literature for various exotic models
of gravity, e.g., the work by Iorio et al.~\cite{Iorio2015a} which can be retrieved from Table
1 of Iorio~\cite{Iorio2015b}. These have been determined by Pitjev and Pitjeva~\cite{Pitjev2013,Pitjeva2013} with the
EPM ephemerides and by Fienga et al.~\cite{Fienga2011} with the INPOP ephemerides. In this
way, it can be discovered from Table 3 that how tremendously outdated is the
uncertainty of 0.21 arcseconds per century as shown in Table 1. Indeed, the latest
uncertainties are much more smaller and Table 3 further gives stringent bound on 
the model parameters.
\begin{table}
\centering 
\caption{Perihilion precession for different parametric values with higher precision} \label{table3}
\begin{tabular}{@{}lrrrrrrrrr@{}}
\hline 
$\lambda_0$  &~~$\beta_0$  &~~$u_0/b^2$  &~~$\Delta \phi $ \\
                   &                   &                     & (arcsec/cen)\\             
\hline 
Venus & 0.009 & 0.1 & 8.736\\
         & 0.01 & 0.1 & 8.630\\
\hline 
Mars & 0.01 & 0.1 & 1.353\\
\hline 
Saturn &  0.009 & 0.099 & 0.0137\\
          & 0.01 & 0.1 & 0.0138\\
\hline 
\end{tabular}
\end{table}

\section*{Acknowledgments} 
FR and SR are grateful to the IUCAA, Pune, India for providing Associateship
Programme. AA is thankful to UGC for providing financial support under the scheme Dr. D.S. Kothari postdoctoral fellowship.  
FR and SM are  also thankful to DST-SERB and CSIR, Government of India for financial support. We all are thankful to the 
anonymous referee for several pertinent suggestions which have enabled us to modify the manuscript substantially.

\end{document}